\documentclass[12pt,nofootinbib,aps,amsmath,amssymb,preprint,showpacs]{revtex4}

\usepackage[T1]{fontenc} \usepackage[latin1]{inputenc}
\usepackage{amsmath,color,ulem} \usepackage{graphicx}
\usepackage{epsf,bm} \usepackage{amssymb}

\usepackage{tabulary}
\newcolumntype{K}[1]{>{\centering\arraybackslash}p{#1}}
\usepackage{multirow}
\begin{document}
 
\title{Approximate dynamical eigenmodes of the Ising model with local
  spin-exchange moves }
\author{Wei Zhong$^\dagger$}
\email{w.zhong1@uu.nl}
\author{Debabrata Panja$^\dagger$}
\author{Gerard T. Barkema$^\dagger$}

\affiliation{
 $^\dagger$Department of Information and Computing Sciences, Utrecht University, Princetonplein 5, 3584 CC Utrecht, The Netherlands\\
}

\date{\today} 

\begin{abstract}
  We establish that the Fourier modes of the magnetisation serve as
  the dynamical eigenmodes for the two-dimensional Ising model at the
  critical temperature with local spin-exchange moves, i.e., Kawasaki
  dynamics. We obtain the dynamical scaling properties for these
  modes, and use them to calculate the time evolution of two dynamical
  quantities for the system, namely the autocorrelation function and
  the mean-square deviation of the line magnetisations. At
  intermediate times $1 \lesssim t \lesssim L^{z_c}$, where $z_c=4-\eta=15/4$
  is the dynamical critical exponent of the model, we find that the
  line magnetisation undergoes anomalous diffusion. Following our
  recent work on anomalous diffusion in spin models, we demonstrate
  that the Generalized Langevin Equation (GLE) with a memory kernel
  consistently describes the anomalous diffusion, verifying the
  corresponding fluctuation-dissipation theorem with the calculation
  of the force autocorrelation function.
\end{abstract}

\maketitle

\section{Introduction \label{Sec1}}
  
For physical systems in statistical physics, the eigenvalues and eigenvectors (of the Hamiltonians) play a central role.
The eigenvectors form a complete orthogonal basis in the
space of variables used to express the Hamiltonian. The eigenvalues
and eigenfunctions identify the ground and the excited states, as well
as their energies, which then form the groundwork for obtaining the
partition function, the principal quantity of interest for calculating
all equilibrium ensemble-averaged observables.

For classical systems, the Hamiltonian also dictates the dynamics of
systems through the equations of motion. Here too, theoretically, the
same concept holds, viz. with the equation of motion of a degree of
freedom $q$ used to describe a Hamiltonian ${\cal H}$ being given by
\begin{eqnarray}
\zeta\dot q=-\frac{\partial {\cal H}}{\partial q},
\label{e0}
\end{eqnarray}
with $\zeta$ being the friction coefficient in the overdamped limit,
it really is an asset to know the {\it dynamical\/} eigenvalues and
eigenvectors. Together, the dynamical eigenvalues and eigenvectors ensure that the full time-dependence of any dynamical quantity
can be calculated exactly.

In contrast to eigenvalues and eigenvectors of the Hamiltonian itself,
the scope for dynamical eigenvalues and eigenvectors is far more
restricted, for the following reason. The eigenvectors $\{r_\alpha\}$
are linear combinations of all the degrees of freedom $\{q_i\}$,
reducing Eq. (\ref{e0}) to the form
\begin{eqnarray}
\zeta_\alpha\dot r_\alpha=-\lambda_\alpha r_\alpha,
\label{e1}
\end{eqnarray}
with $\lambda_\alpha$ being the corresponding {\it dynamical\/}
eigenvalue, obtained from the diagonalisation of the Hessian matrix
$\displaystyle{\frac{\partial^2{\cal H}}{\partial r_\beta\partial
    r_\gamma}}$. The dynamical eigenmodes $\{r_\alpha\}$, if they
exist, are often simply called the {\it modes\/} of the system. For
the form (\ref{e1}) to hold, the Hessian must be independent of
$\{r_\alpha\}$, which restricts the class of such Hamiltonians only to
harmonic ones (i.e., ${\cal H}$ is quadratic in $\{q_i\}$). Classic
examples of such systems are the bead-spring models of linear
polymeric systems \cite{doi1,doi2}, their extensions to star and
tadpole polymers \cite{rick1}, polymeric membranes
\cite{nelson,rick2}, 2D cytoskeleton of cells \cite{lipo,picart,sack}
and graphite oxide sheets \cite{sack,hwa,spec,wen}.

Not all is however lost if the Hamiltonian is not harmonic (which is
in fact almost always the case). Note here that {\it any\/} complete
orthogonal basis in the space of the degrees of freedom can be used to
describe the dynamics of the system. The main disadvantage of choosing
an arbitrary one is that the corresponding amplitudes remain
dynamically (nonlinearly) coupled at all times, preventing one from
taking large time-steps in computer simulations. Despite this
shortcoming, sometimes one can be lucky to realise that there are {\it
  approximate\/} modes that can allow one to take somewhat large
time-steps within a preordained error margin. Examples are the Rouse
modes for self-avoiding polymers \cite{panja1}, a reptating polymer
chain \cite{panja2}, and polymer chains in a melt
\cite{kalathi1,kreer,kalathi2}.

The focus of the present paper are the (approximate dynamical) modes
of the two-dimensional (2D) square-lattice Ising model (system size
$L\times L$) with local spin exchange moves --- commonly known as
Kawasaki moves \cite{kawasaki} --- at critical temperature and at zero
order parameter, introduced in Sec. \ref{Sec2.1}. We focus on the line
magnetisation for this model and find, surprisingly, that the Fourier
modes provide a very good approximation of the true dynamical
eigenmodes. We numerically investigate the properties of these modes
in Sec. \ref{Sec2.2}-\ref{Sec2.4}, numerically revealing that the
equilibrium amplitude of the $p$-th mode behaves as
$ (L/p)^{\gamma/\nu}\,(B_0+B_1 p^{-\gamma/\nu})$, and that its decay time scales
$\sim(L/p)^{z_c} $, where $\gamma=7/4$, $\nu=1$ and
$\eta=2-\gamma/\nu=1/4$ are the three equilibrium critical exponents
of the Ising model, and $z_c=4-\eta=15/4$ is the critical dynamical
exponent for the model with local spin exchange moves
\cite{halp,yala,alex}. In Sec. \ref{Sec3} we use these results to
analytically calculate two observables: the autocorrelation function,
and the mean-square deviation (MSD), of the line magnetisation. We
find that line magnetisation exhibits anomalous diffusion. Our results
for anomalous diffusion is consistent with a pattern that the dynamics
of magnetisation at the critical temperature in spin models is
anomalous \cite{walter,zhong1,zhong2}. Importantly, the anomalous
diffusion is described by the Generalised Langevin Equation (GLE)
\cite{zhong1,zhong2} (and bears strong resemblance to anomalous
diffusion in polymeric and membrane systems under a variety of
circumstances
\cite{rick1,rick2,panja1,panja1a,panja1b,panja2b,maes,kroy,popova,mizuochi,panja3a,panja3b,panja3c,dubbel,panja4,panja5,sakaue}),
which we verify in Sec. \ref{Sec4}. We conclude the paper in
Sec. \ref{Sec5}.
 
\section{The model and the Fourier modes as the approximate dynamical modes\label{Sec2}}

\subsection{Ising model with local spin-exchange (Kawasaki) dynamics\label{Sec2.1}}

We consider the two-dimensional (2D) Ising model on an $L\times L$ square
lattice with periodic boundary conditions in both $x$- and
$y$-directions. The Hamiltonian for the model is given by
 \begin{equation}
{\cal H}=-J\sum_{\langle (j,k)(m,n) \rangle}s_{j,k}\,\,s_{m,n},
\label{eq1.1}
\end{equation}
where $s_{j,k} = \pm1$ is the spin value at $x$-location $j$ and
$y$-location $k$, and $J$ is the coupling constant for interactions
among the spins and we set $J=1$ during our simulations. The summation runs over all the nearest-neighbour
spins, and $0 \leq (j,k,m,n) < L$. All properties we report here have
been obtained by simulating the model at the critical temperature
$T_c=2/\ln(1+\sqrt2)$, and by setting the value of
the Boltzmann constant $k_B$ to unity.

The model is simulated with Kawasaki dynamics at $T_c$. All
simulations reported in this paper have been performed at zero
(conserved) order parameter. In other words, we fix the total
magnetisation of the system at zero, and at each Monte Carlo move, two
neighbouring spins are randomly selected to exchange their values. The
resulting energy change $\Delta E$ is measured, and the move is
accepted with the normal Metropolis probability min$[1,\exp(-\Delta
  E/T)]$. For each unit of time, on average, all the spins are supposed to be
selected once.
 
\subsection{Fourier modes for line magnetisation\label{Sec2.2}}

In this model we define the line magnetisation as
$\displaystyle{M_l(j,t)=\sum_{k=0}^{L-1}s_{j,k}}(t)$; correspondingly,
the $p$-th Fourier mode amplitude of the line magnetisation is given
by
\begin{equation}
A_p(t)=\frac{1}{L}\sum_{j=0}^{L-1}M_l(j,t)\exp(-2\pi i pj/L)=X_p(t)-iY_p(t),
\label{eq1.2}
\end{equation} 
where   
\begin{eqnarray}
X_p(t)&=&\frac{1}{L}\sum_{j=0}^{L-1}M_l(j,t)\cos(2\pi pj/L),
\nonumber\\
Y_p(t)&=&\frac{1}{L}\sum_{j=0}^{L-1}M_l(j,t)\sin(2\pi pj/L),
\label{eq1.4}
\end{eqnarray} 
respectively are the real and the imaginary parts of the Fourier
transform, with $p=0, 1,\dots,(L-1)$. The inverse Fourier transform is
then given by
\begin{eqnarray}
M_l(j,t)&=&\sum_{p=0}^{L-1}A_p(t)\exp(2\pi i pj/L)\quad\text{or}
\nonumber\\
M_l(j,t)&=&\sum_{p=0}^{L-1} \left[X_p(t)\cos(2\pi pj/L)+Y_p(t)\sin(2\pi pj/L)\right].
\label{eq1.6}
\end{eqnarray}

\subsection{Equilibrium properties of the Fourier mode amplitudes \label{Sec2.3}}

We express the equilibrium correlations of the Fourier modes as
\begin{equation}
X_{pq}(t)=\langle X_p(t)X_q(0)\rangle\quad\text{and}\quad Y_{pq}(t)=\langle Y_p(t)Y_q(0)\rangle,
\end{equation}
where the angular brackets ($\langle\cdot\rangle$) define an average
over equilibrated ensembles.

The cross-correlation terms, $\langle X_p(t) Y_q(0)\rangle$ and
$\langle Y_p(t) X_q(0)\rangle$ respectively, can be argued to be equal
to zero, as follows. Let us consider $\langle X_p(t) Y_q(0)\rangle$ to
illustrate the calculation. First, having expressed it as
$\displaystyle{\sum_{j,m=0}^{L-1}\left\langle
  M_l(j,0)M_l(m,t)\right\rangle\cos(2\pi pj/L)\sin(2\pi qm/L)}$, then
making the simultaneous substitutions $j\rightarrow (L-j)$ and
$m\rightarrow (L-m)$, and finally using $M_l(0,t)=M_l(L,t)$ due to
periodic boundary conditions, we find that the term also equals
$-\displaystyle{\sum_{j,m=0}^{L-1}\left\langle
  M_l(L-j,0)M_l(L-m,t)\right\rangle\cos(2\pi pj/L)\sin(2\pi
  qm/L)}$. Next, we use the fact that $\left\langle
M_l(j,0)M_l(m,t)\right\rangle$ is only a function of $|j-m|$ modulo
$L/2$ (due to periodic boundary conditions) as well as only of $|t|$
(due to time reversibility invariance at equilibrium). This implies
that $\left\langle M_l(j,0)M_l(m,t)\right\rangle=\left\langle
M_l(L-j,0)M_l(L-m,t)\right\rangle$, leading to the condition $\langle
X_p(t) Y_q(0)\rangle=-\langle X_p(t) Y_q(0)\rangle=0$. For this reason
we leave both $\langle X_p(t) Y_q(0)\rangle$ and $\langle Y_p(t)
X_q(0)\rangle$ out of further considerations.

\begin{figure}[h]
  \includegraphics[width=0.45\linewidth]{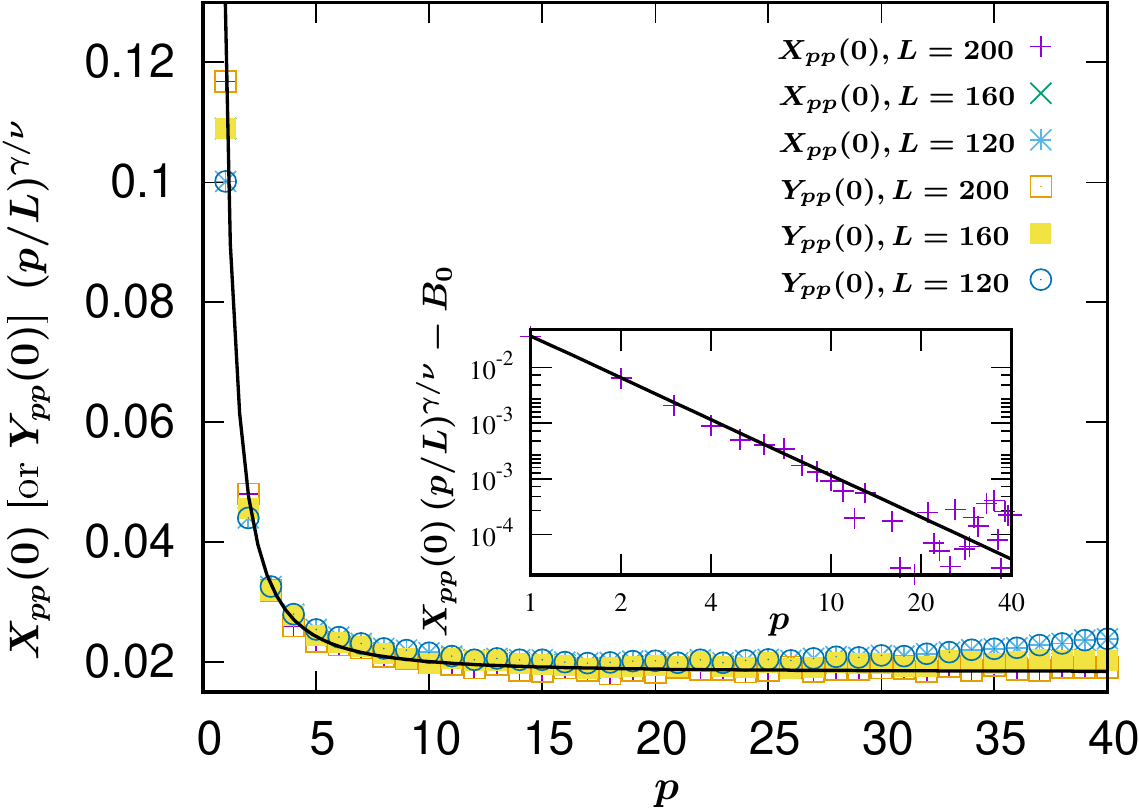}
  \caption{(color online) $X_{pp}(0)$ and $Y_{pp}(0)$ as
    a function of $p$ for different system sizes, with $p=1$ to $40$,
    and $L=120$, $160$, $200$. Fitting to the data leads to
    $X_{pp}(0)=Y_{pp}(0)\approx (L/p)^{\gamma/\nu}\,(B_0+B_1 p^{-\gamma/\nu})$, where $B_0=0.0185$ and $B_1=0.1$. Inset:
    $X_{pp}(0)$ data for $L=200$ is fitted in a log-log plot; the
    straight line has slope $-1.75 (=-\gamma/\nu)$.}
\label{fourier_modes}
\end{figure}

Next, we argue that $X_{pp}(0)=Y_{pp}(0)$ at least up to
$O(L^{-2})$. In order to do so, we first express $Y_{pp}(0)$ as
$\displaystyle{\sum_{j,m=0}^{L-1}\left\langle
  M_l(j,0)M_l(m,0)\right\rangle\sin(2\pi pj/L)\sin(2\pi
  pm/L)}=\displaystyle{\sum_{j,m=0}^{L-1}\left\langle
  M_l(j,0)M_l(m,0)\right\rangle\cos(2\pi pj'/L)\cos(2\pi pm'/L)}$,
where
$(j',m')=\displaystyle{\left(j+\frac{L}{4p},m+\frac{L}{4p}\right)}$. We
then again observe, just like in the above paragraph, that
$\left\langle M_l(j,0)M_l(m,t)\right\rangle$ is only a function of
$|j-m|$ modulo $L/2$. This implies that if
$\displaystyle{\frac{L}{4p}}$ is an integer, then upon relabelling the
line indices the sum trivially reduces to
$\displaystyle{\sum_{j',m'=0}^{L-1}\left\langle
  M_l(j',0)M_l(m',0)\right\rangle\cos(2\pi pj'/L)\cos(2\pi
  pm'/L)}=X_{pp}(0)$. If however $\displaystyle{\frac{L}{4p}}$ is not
an integer, then, we can still relabel the indices as
$\displaystyle{\sum_{j'',m''=0}^{L-1}\left\langle
  M_l(j'',0)M_l(m'',0)\right\rangle\cos(2\pi p(j''+\Delta
  x)/L)\cos(2\pi p(m''+\Delta x)/L)}$, with $\Delta x<1$, 1 being the
lattice unit. Beyond this point, we can do a Taylor expansion of the
cosine terms, implying that the equality $Y_{pp}(0)=X_{pp}(0)$ must
hold up to $O(L^{-2})$. This, together with the scaling of
$\langle|A_p^2|\rangle\sim(L/p)^{\gamma/\nu}$ in the limit
$p\rightarrow\infty$ for the 2D Ising model as derived in Appendix A,
we attempt to fit $X_{pp}(0)=Y_{pp}(0)$ to the asymptotic scaling
$\sim (L/p)^{\gamma/\nu}$ in Fig. \ref{fourier_modes}.

From this fit, we find that $X_{pp}(0)\approx Y_{pp}(0)\approx
(L/p)^{\gamma/\nu}\,(B_0+B_1 p^{-\gamma/\nu}) $, where $B_0=0.0185$ and
$B_1=0.1$ are two numerically obtained constants. Note also that
\begin{equation}
\begin{split}
X_{p(L-q)}(t)=X_{pq}(t)\quad\mbox{and}\quad Y_{p(L-q)}(t)=Y_{pq}(t),
\end{split}
\label{eq1.8}
\end{equation}
an obvious result obtained from the symmetry of the mode amplitudes
under $p\leftrightarrow L-p$.

\begin{figure*}
\includegraphics[width=0.45\linewidth]{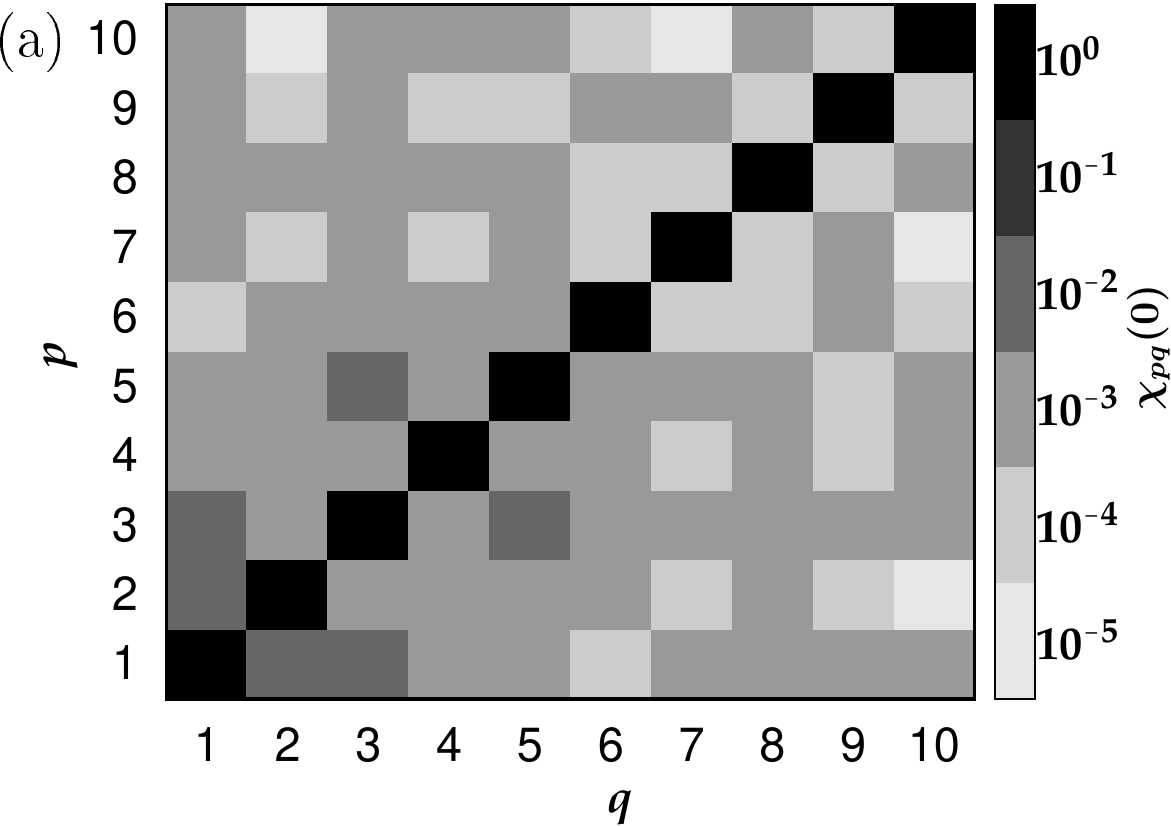}
\hspace{3mm}
\includegraphics[width=0.45\linewidth]{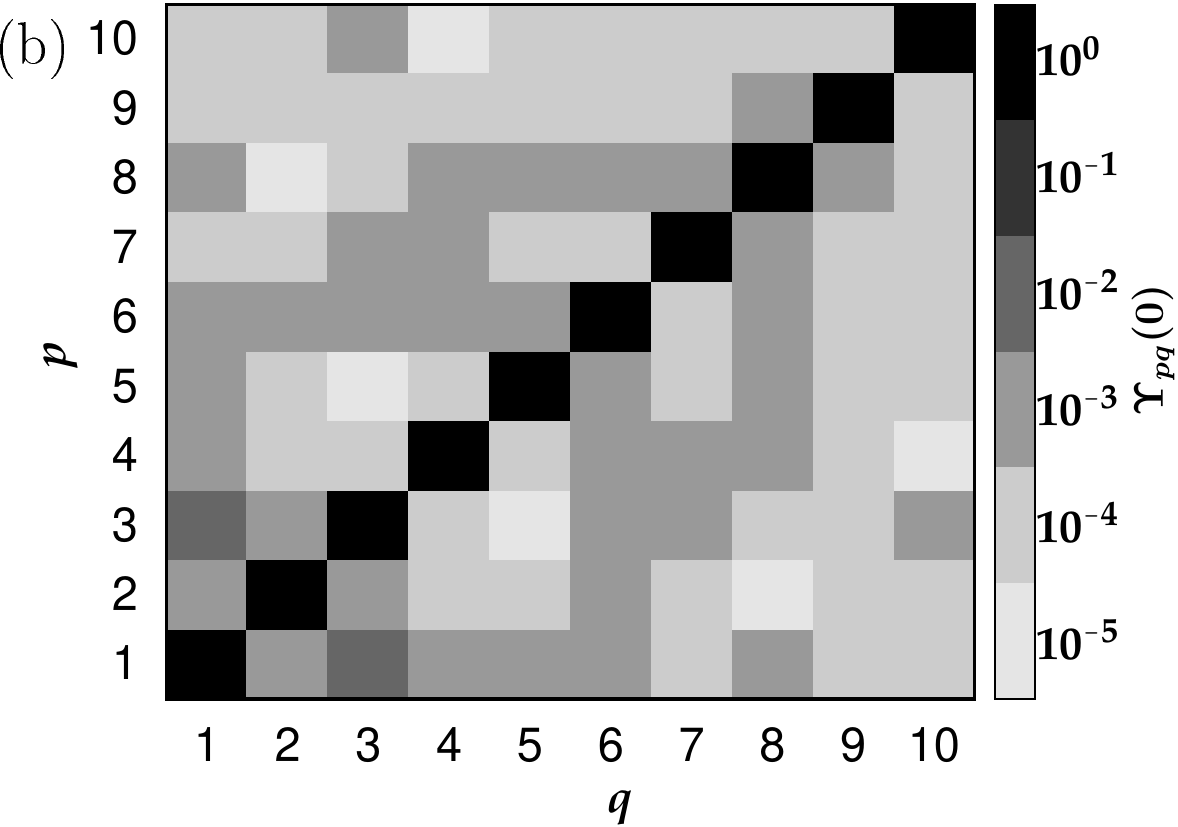}
  \caption{ The matrix (a), $\chi_{pq}(0)\equiv
    X_{pq}(0)/\sqrt{X_{pp}(0)X_{qq}(0)}$ and (b),
    $\Upsilon_{pq}(0)\equiv Y_{pq}(0)/\sqrt{Y_{pp}(0)Y_{qq}(0)}$ in
    logarithmic greyscale for $p,q=1,2,...,10$ and $L=40$. The values
    of the off-diagonal elements of $\chi_{pq}(0)$ and
    $\Upsilon_{pq}(0)$ are not zero. However, most of them are
    typically two or more orders of magnitude smaller than the
    diagonal ones, which means the modes are statistically uncorrelated.}
\label{matrix}
\end{figure*}

The results of Fig. \ref{fourier_modes} are supplemented with the data
for $\chi_{pq}(0)\equiv X_{pq}(0)/\sqrt{X_{pp}(0)X_{qq}(0)}$ and
$\Upsilon_{pq}(0)\equiv Y_{pq}(0)/\sqrt{Y_{pp}(0)Y_{qq}(0)}$ for
$L=40$ and $p,q<L/2$ (specifically, $p,q=1$ to $10$) in
Fig. \ref{matrix}. The values of the off-diagonal elements of
$\chi_{pq}(0)$ and $\Upsilon_{pq}(0)$ are not zero (we do not expect
them to be zero even after caring for numerical accuracy); however,
they are at least two orders of magnitude smaller than the diagonal
ones.

Together these results indicate that to a very good approximation the
modes remain statistically independent during the system's evolution
by means of Kawasaki dynamics.

\subsection{Fourier modes as approximate dynamical eigenmodes of the model\label{Sec2.4}}

\begin{figure*}
\includegraphics[width=0.45\linewidth]{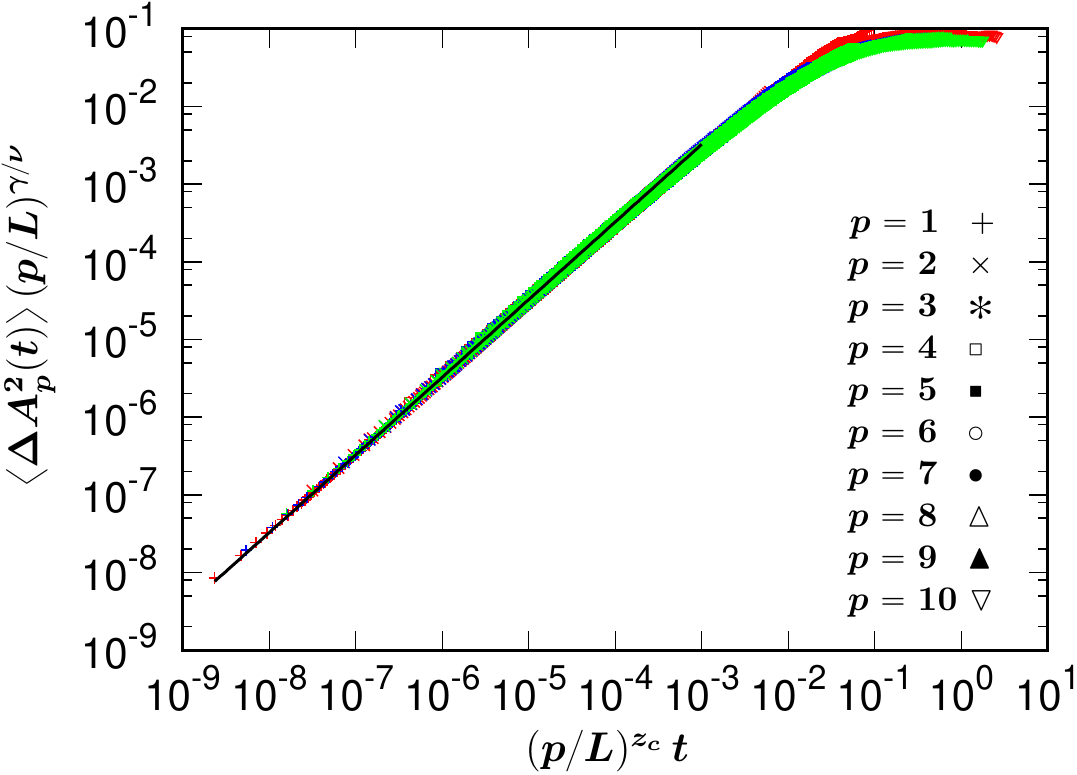}
  \caption{(color online) The MSD of the complex modes amplitude
    $\langle|\Delta A_p^2(t)|\rangle=\sqrt{2}\,\langle\Delta Y_p^2(t)
    \rangle=\sqrt{2}\,\langle\Delta X_p^2(t) \rangle$, where
    $\langle\Delta X_p^2(t) \rangle=\langle [X_p(t)-X_p(0)]^2\rangle$
    and $\langle\Delta Y_p^2(t) \rangle=\langle
    [Y_p(t)-Y_p(0)]^2\rangle$. For every system size $L=\,$120 (red),
    160 (blue), 200 (green), the MSD of ten different mode amplitudes
    are measured. In the range $t \lesssim (p/L)^{z_c}$, the modes
    shows normal diffusion and the solid line represents $\langle|\Delta
    A_p^2(t)|\rangle \, (p/L)^{\gamma/\nu}\approx
    3.2527\,(p/L)^{z_c}\,t$.}
\label{modes_msd}
\end{figure*}

In Fig. \ref{modes_msd}(a), we obtain a data collapse plot for the
mean-square deviation (MSD) of the complex mode amplitude
$\langle|\Delta A_p^2(t)|\rangle$, as a function of $(p/L)^{z_c}t$ for
$p=1,2,\ldots,10$ for three different system sizes $L=120,160,200$
(from our earlier works on spin systems \cite{walter,zhong1,zhong2} we
expect that the data collapse would require scaling time with a
prefactor $(p/L)^{z_c}$). The solid line in the figure then represents
\begin{eqnarray}
\langle\Delta A_p^2(t) \rangle= \sqrt{2} \langle\Delta X_p^2(t) \rangle=\sqrt{2} \langle\Delta Y_p^2(t)
  \rangle\approx 3.2527 \,\displaystyle{\left( \frac{L}{p}\right)^{\gamma/\nu}}(p/L)^{z_c} t\quad \mbox{for} \quad (p/L)^{z_c} t\ll1.
  \label{msd_res}
\end{eqnarray}

Since the MSDs of the mode amplitudes can be expressed in terms of
their autocorrelation functions as
\begin{eqnarray}
  \langle\Delta X_p^2(t) \rangle &= &\langle [X_p(t)-X_p(0)]^2\rangle=2X_{pp}(0)\left[1-\frac{X_{pp}(t)}{X_{pp}(0)}\right]
    \nonumber\\
 \langle\Delta Y_p^2(t) \rangle &=&\langle [Y_p(t)-Y_p(0)]^2\rangle=2Y_{pp}(0)\left[1-\frac{Y_{pp}(t)}{Y_{pp}(0)}\right],
 \label{msd1}
\end{eqnarray}
with the approximation
$X_{pp}(0)=Y_{pp}(0)\approx (L/p)^{\gamma/\nu}\,(B_0+B_1 p^{-\gamma/\nu})$,
for $(p/L)^{z_c} t\ll1$ in a large range shown in
Fig. \ref{modes_msd}, Eqs. (\ref{msd_res}-\ref{msd1}) can be recast in
the form
\begin{equation}
\frac{X_{pp}(t)}{X_{pp}(0)}=\frac{Y_{pp}(t)}{Y_{pp}(0)}\approx \exp \left[-\frac{1.15 (p/L)^{z_c} t}{0.0185+0.1\, p^{-\gamma/\nu}}\right].
\label{xyppt}
\end{equation}  

To conclude, in this section we have demonstrated that to a very good
approximation the Fourier modes for the 2D Ising model with Kawasaki
dynamics remain statistically uncorrelated at all times, and their
autocorrelations decay exponentially in time, from which we conclude
that they are approximate dynamical eigenmodes. This means that the
properties of the modes amplitude can be used to calculate all
dynamical quantities to a very good approximation
\cite{doi1,doi2,panja1,rick1}. In the following section, we will
showcase this to calculate the autocorrelation function and the MSD of
line magnetisations.

\section{Dynamics of two physical observables using the Fourier modes as approximate dynamical eigenmodes\label{Sec3}}

In this section we focus on the dynamics observables of the
system. Using the properties of the Fourier modes obtained in the last
section, we analytically derive the autocorrelation function and the
mean-square deviation of the line magnetization.
 
\subsection{Autocorrelation function of the line magnetisation \label{Sec3.1}}

The first dynamical observable we are dealing with is the
autocorrelation function of the line magnetisation, defined
as
\begin{equation}
  C(t)=\langle M_l(x,t)M_l(x,0)\rangle.
  \label{eq3.1}
\end{equation}
\begin{figure}[!h]
 \includegraphics[width=0.5\linewidth]{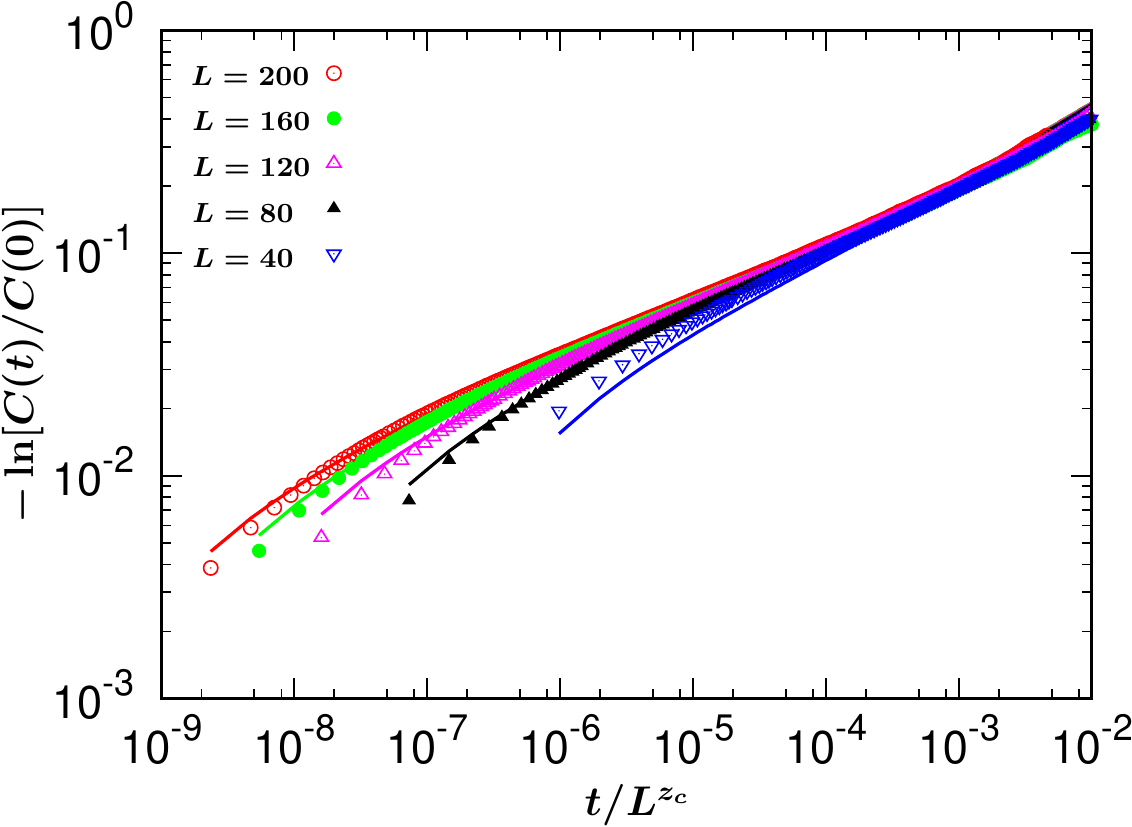}
  \caption{(color online) Comparison between the simulation results
    (points) and expectation values from Eq. (\ref{eq3.2}) (solid
    lines, same colours as the points) for the autocorrelation
    function $C(t)$ of the line magnetisation, for different system
    sizes.}
\label{autocorr}
\end{figure}

This autocorrelation function can be expressed in terms of the modes
by combining Eqs. (\ref{eq1.6}), (\ref{eq1.8}) and (\ref{eq3.1}),
yielding
\begin{equation}
  \begin{split}
 C(t) &=4\sum_{p=1}^{L/2} X_{pp}(t) \\
 &= 4\,\sum_{p=1}^{L/2}\left(\frac Lp\right)^{\gamma/\nu}\exp\left[-\frac{1.15 \,(p/L)^{z_c} t}{0.0185+0.1\,p^{-\gamma/\nu}}\right] \left(0.0185+0.1p^{-\gamma/\nu}\right)\\
 \end{split}
  \label{eq3.2}
 \end{equation}
As shown in Fig. \ref{autocorr}, the prediction (\ref{eq3.2}) fits the
simulation results quite well.
 
\subsection{Anomalous diffusion of the line magnetisation \label{Sec3.2}}

Let us now consider the MSD of the line magnetisation 
\begin{equation}
\langle \Delta M_l^2(t)\rangle=\langle [M_l(x,t)-M_l(x,0)]^2\rangle
\label{eq4.1}
\end{equation}
as another dynamical observable.

Using Eq. (\ref{eq1.6}) and $\langle X_p(t)Y_q(0)\rangle=\langle
Y_p(t)X_q(0)\rangle=0$, we have
\begin{equation}
\begin{split}
\langle \Delta M_l^2(t)\rangle=&\sum_{p=0}^{L-1} \sum_{q=0}^{L-1}\langle [X_p(t)-X_p(0)][X_q(t)-X_q(0)]\cos(2\pi px/L)\cos(2\pi qx/L)\\
 &+[Y_p(t)-Y_p(0)][Y_q(t)-Y_q(0)]\sin(2\pi px/L) \sin(2\pi qx/L)\rangle.
\end{split}
\label{eq4.2}
\end{equation}
 
Then Eq. (\ref{eq4.2}) can be simplified with the approximation
$X_{pq}(t)=Y_{pq}(t)=X_{p (L-q)}(t)=Y_{p (L-q)}(t)$, and $X_0(t)$ as
the conserved order parameter (chosen to be zero) of the dynamics,
leading us to
\begin{equation}
\begin{split}
\langle \Delta M_l^2(t)\rangle &=2\,\sum_{p=1}^{L-1} \sum_{q=1}^{L-1}[X_{pq}(0)-X_{pq}(t)]\\
&=8\,\sum_{p=1}^{L/2} \sum_{q=1}^{L/2}X_{pq}(0)\left[1-\frac{X_{pq}(t)}{X_{pq}(0)}\right]\\
&=8\,\sum_{p=1}^{L/2} X_{pp}(0)\left[1-\frac{X_{pp}(t)}{X_{pp}(0)}\right].
\end{split}
\label{eq4.3}
\end{equation}

Using the properties of $\displaystyle{X_{pp}(t)}$ and $\displaystyle{X_{pp}(0)}$
as obtained in Secs. \ref{Sec2.3}-\ref{Sec2.4}, the behavior of the
MSD of the line magnetisation can be divided into two time
domains.

\begin{figure}[h]
\includegraphics[width=0.45\linewidth]{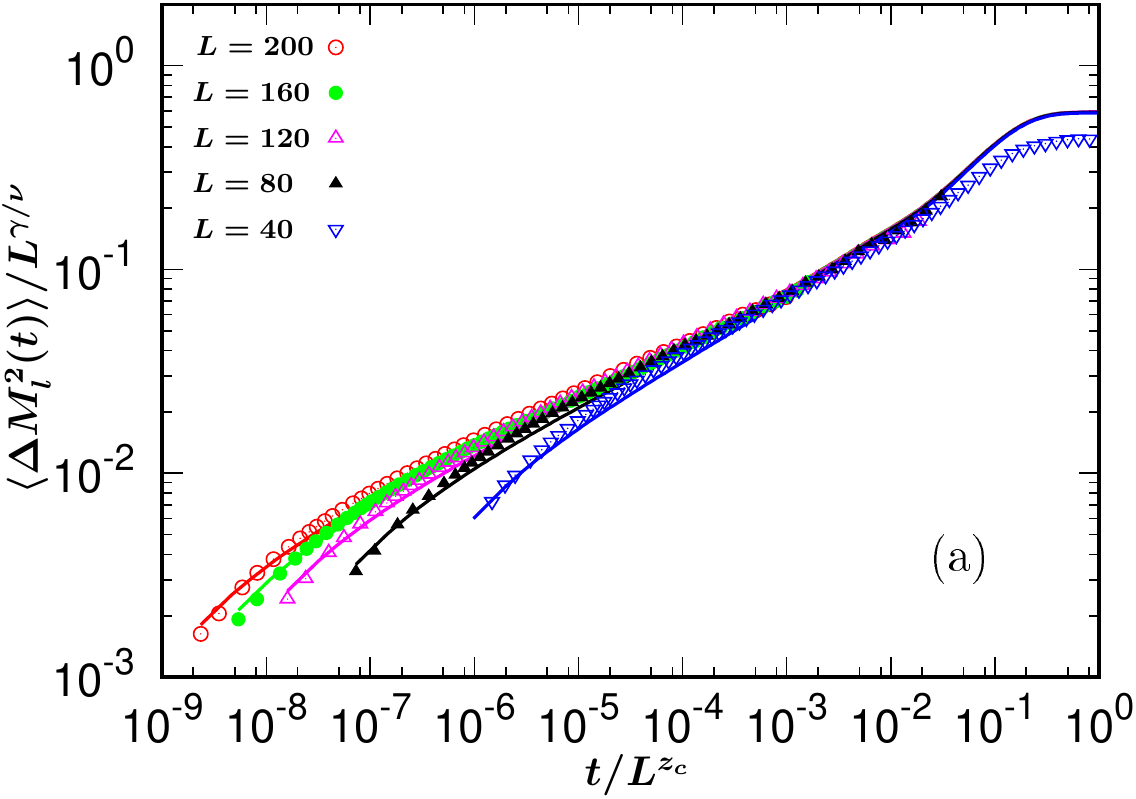}\hspace{1cm}\includegraphics[width=0.43\linewidth]{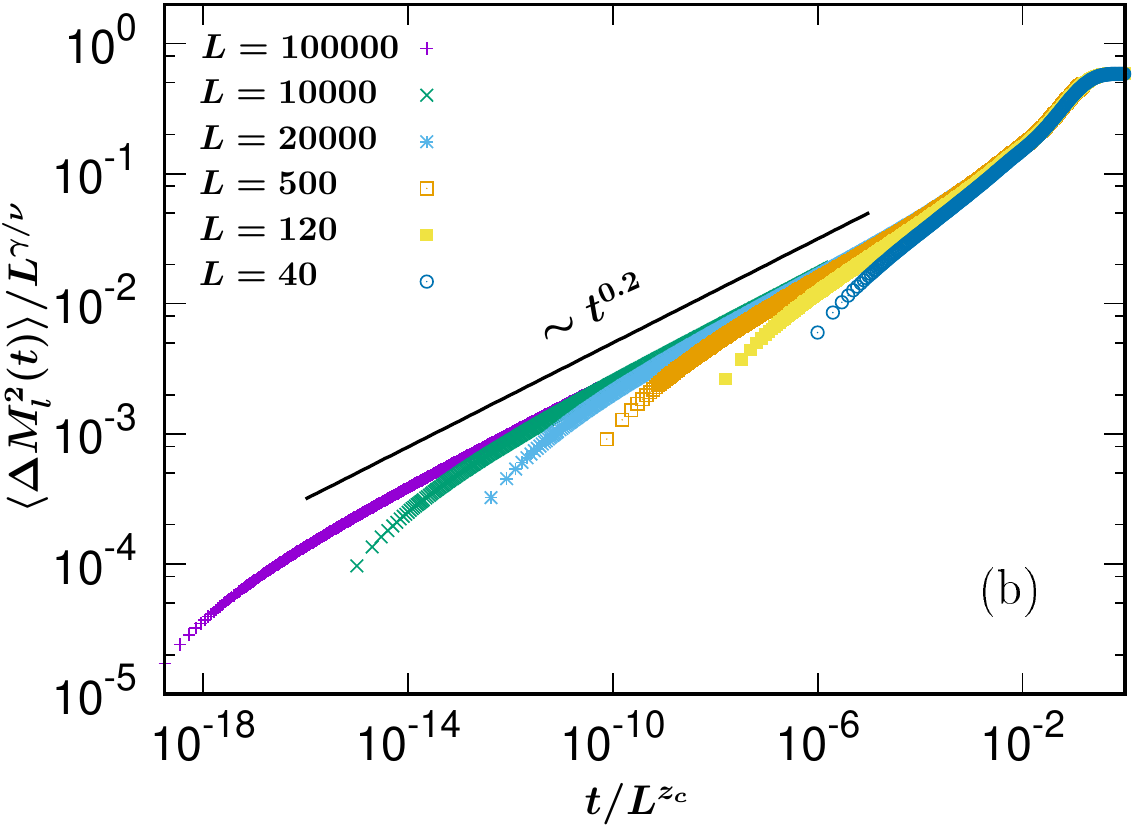}
  \caption{(color online) (a) Comparison between the simulation
    results (points) and the results obtained from Eq. (\ref{eq4.5})
    (solid lines, same colour as the points) for the MSD of the line
    magnetisation $\langle\Delta M_l^2(t) \rangle$ for different
    system sizes. (b) Confirmation of the sum (\ref{eq4.5b}) to
    power-law $t^{(\gamma/\nu-1)/z_c}\approx t^{0.2}$ for $L\rightarrow\infty$.}
\label{msd}
\end{figure}

At long times $t\gtrsim L^{z_c}$,
$\displaystyle{\frac{X_{pp}(t)}{X_{pp}(0)}}\rightarrow0$, meaning that
$\langle\Delta M_l^2(t)\rangle$ approaches a constant $\sim
L^{\gamma/\nu}$. At intermediate times $1\lesssim t\lesssim L^{z_c}$,
\begin{equation}
\begin{split}
 \langle \Delta M_l^2(t)\rangle\! &= \!
 8\,\sum_{p=1}^{L/2}\,
 X_{pp}(0)\!\left[1\!-\!\frac{X_{pp}(t)}{X_{pp}(0)}\right]\!  \\ 
 &=\!
 8\,\sum_{p=1}^{L/2} \left(\frac Lp\right)^{\gamma/\nu} \left[1\!-\!\exp\left(-\frac{1.15\,(p/L)^{z_c} t}{0.0185+0.1\,p^{-\gamma/\nu}}\right)\right] \left(0.0185+0.1p^{-\gamma/\nu}\right).
\end{split}
  \label{eq4.5}
\end{equation}
As shown in Fig. \ref{msd} (a), the prediction (\ref{eq4.5}) fits the
simulation results quite well.

For an analytical expression for the msd, with $x=p/L$, the sum
(\ref{eq4.5}) can be reduced to the following integral:
\begin{equation}
\begin{split}
\langle \Delta M_l^2(t)\rangle =8L\!\!\int_{1/L}^{1/2}\!\!\frac{dx}{x^{\gamma/\nu}}\left(1\!-\!\exp\left[-\frac{1.15\,tx^{z_c} }{\{0.0185+0.1/(xL)^{\gamma/\nu}\}}\right]\right) \left\{0.0185+0.1/(xL)^{\gamma/\nu}\right\},
\end{split}
  \label{eq4.5a}
\end{equation}
but beyond that it is difficult process it further without making
approximations. In particular, in the limit $L\rightarrow\infty$ and
finite values of $x$, the second term within the curly brackets can be
dropped. At the lower limit of $x$, the two terms within the curly
brackets are however comparable. Nevertheless, if we do drop this
second term altogether, then the integral can be easily performed to
show that in the leading order of $L$
\begin{equation}
\begin{split}
\langle \Delta M_l^2(t)\rangle\sim L\left(\frac tL\right)^{(\gamma/\nu-1)/z_c} \Rightarrow \langle \Delta M_l^2(t)\rangle\sim L^{\gamma/\nu}\left(\frac{t}{L^{z_c}}\right)^{(\gamma/\nu-1)/z_c}\approx L^{\gamma/\nu}\left(\frac{t}{L^{z_c}}\right)^{0.2}.
\end{split}
  \label{eq4.5b}
\end{equation}
This behaviour of the sum (\ref{eq4.5b}) is shown in Fig. \ref{msd}(b).
 
\section{Generalised Langevin Equation formulation for the anomalous diffusion in the Ising model with Kawasaki dynamics \label{Sec4}}

In Sec. \ref{Sec3} we have demonstrated that at the intermediate time
regime, the line magnetisation in the Ising model with Kawasaki
dynamics exhibits anomalous diffusion. In our recent studies on the
Ising and $\phi^4$ model with Glauber dynamics \cite{zhong1,zhong2},
we have argued that the anomalous diffusion of the magnetization
belongs to the GLE class, for which the restoring force plays an
important role.

Imagine that we choose a tagged line, and since the thermal spin
flips, at $t=0$ its magnetisation $M_l$ changes by a little amount
$\delta\!M_l$. The surrounding spins will react to this change due to
the interactions dictated by the Hamiltonian, and it takes time to
spread this reaction. During this time, the value of $M_l$ will also
readjust to the persisting values of the surrounding spins, undoing at
least a part of $\delta\!M_l$. It is the latter that we interpret as
the result of ``inertia'' of the surrounding spins that resists
changes in $M_l$, and the resistance itself acts as the restoring
force to the changes in the tagged magnetisation, and finally, leads
to anomalous diffusion.

\subsection{Generalized Langevin Equation for the line magnetisation \label{sec4.1} }

From how the restoring force works introduced before, it not only
indicates that there is a memory effect which is significant during
the `restoring' process, but also leads us to the GLE formulation to
describe the anomalous diffusion.

In line with our previous works on the Ising and $\phi^4$ model with
Glauber dynamics \cite{zhong1,zhong2} and in
polymeric systems \cite{rick1,panja1a,panja1b,panja2b}, the relation
of the restoring force $f(t)$ and the ``velocity'' of magnetisation
$\dot M_l(t)$ can be expressed as
\begin{subequations}
\begin{equation}
\zeta \dot{M_l}(t) = f(t)+q_1(t)
 \label{c2_1}
\end{equation}   
\begin{equation}
f(t) = -\int_0^tdt'\mu(t-t')\,\dot{M_l}(t')+q_2(t).
 \label{c2_2}
\end{equation}
 \label{c2}
\end{subequations}
Here $f(t)$ is the internal force, $\zeta$ is the ``viscous drag'' on
$M_l$, $\mu(t-t')$ is the memory kernel, $q_1(t)$ and $q_2(t)$ are
two noise terms satisfying $\langle q_1(t)\rangle=\langle
q_2(t)\rangle=0$, and the fluctuation-dissipation theorems (FDTs) are given by
$\langle q_1(t)\,q_1(t')\rangle\propto\zeta\delta(t-t')$ and $\langle
q_2(t)\,q_2(t')\rangle\propto\mu(t-t')$ respectively.

Equation (\ref{c2_2}) can be inverted to write as
\begin{equation}
\dot M_l(t)=-\int_0^t dt'\,a(t-t') f(t')+\omega(t).
\label{c3}
\end{equation}
The noise term $\omega(t)$ similarly satisfies
$\langle\omega(t)\rangle=0$, and the FDT $\langle\omega(t)
\omega(t')\rangle=a(|t-t'|)$. Then $a(t)$ and $\mu(t)$ are related to
each other in the Laplace space as $\tilde a(s)\tilde\mu(s)=1$.
 
To combine Eq. (\ref{c2_1}) and (\ref{c2_2}), we obtain
\begin{equation}
\zeta \dot{M_l}(t)=-\int_0^tdt'\mu(t-t')\,\dot{M_l}(t')+q_1(t)+q_2(t).
\label{c4}
\end{equation}  
or
\begin{equation}
\dot{M_l}(t)=-\int_0^tdt'\theta(t-t')\,[q_1(t)+q_2(t)].
\label{c5}
\end{equation} 
where in the Laplace space $\tilde
\theta(s)[\zeta+\tilde\mu(s)]=1$. With $t > t'$, without any loss of
generality, using Eq. (\ref{c5}) the result of the velocity
autocorrelation is
\begin{equation}
\langle \dot{M_l}(t) \dot{M_l}(0)\rangle\sim \theta(t-t'),
\label{c6}
\end{equation}
 where $\theta(t)$ can be calculated by Laplace inverting the relation
 $\tilde \theta(s)[\zeta+\tilde\mu(s)]=1$.
 
If the memory term is a power law in time, i.e.,
\begin{equation}
 \mu(t)\sim t^{-c}.
\end{equation}

Using the results from Ref. \cite{panja1b}, we have
\begin{equation}
\langle \dot{M_l}(t) \dot{M_l}(0) \rangle|_{f=0}\sim -(t-t')^{c-2}.
\label{vel_auto}
\end{equation}
 
 By integrating Eq. (\ref{vel_auto}) twice in time, we obtain that 
 \begin{equation}
 \langle \Delta M_l ^2(t)\rangle \sim t^c.
 \end{equation}
 
In summary, there is a power-law memory function $\mu(t)\sim t^{-c}$ which
plays a vital part in the GLE formulation. From this we can deduce
that the anomalous diffusion found in Eq. (\ref{eq4.5}) is
non-Markovian and the anomalous exponent is $c$.
 
\subsection{Verification of the power-law behavior of $\mu(t)$ \label{Sec4.2}} 
 
Based on the FDT mentioned under Eq. (\ref{c2_2}), we now numerically
verify the behavior of $\mu(t)$.
 
During simulations, at $t=0$, we thermalise the system to its
equilibrium state. For $t>0$ we select a line and fix its value of the
magnetisation $M_l$ by performing non-local spin-exchange dynamics,
i.e., we choose two lattice site $(j,k)$ and $(m,n)$, if $s_{j,k}
s_{m,n}=-1$ then we exchange their values, else we keep their values
as they are. The energy change $\Delta E$ is measured and we accept
the move with the Metropolis probability $min(1,\exp(-\Delta
E/T))$. For the rest of the system, we let them evolve with the
Kawasaki dynamics.
\begin{figure}[h]
\includegraphics[width=0.45 \linewidth]{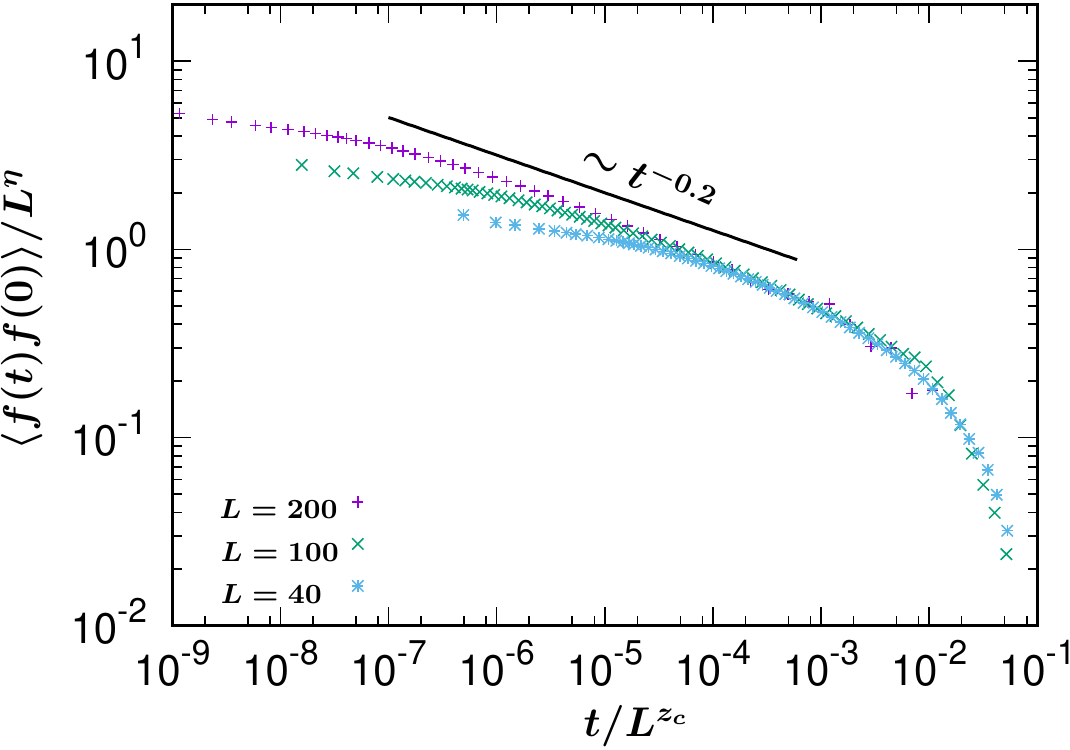}
  \caption{(color online) The autocorrelation function $\langle
    f(t)f(0)\rangle$ as a function of time; the solid line corresponds
    to $\langle f(t)f(0)\rangle\sim t^{-(\gamma/\nu-1)/z_c}\approx t^{-0.2}$. }
  \label{fdt}
\end{figure}
  
We then keep taking snapshots of the system at regular intervals. For
every snapshot we take, we consider an attempt to flip each spin in
turn and find the expected change in $M_l$ which would have occurred
if this move had been implemented, totalled over all the spins on the
selected line, and the possible change of the line magnetisation is
defined as $f(t)=\dot{M}(t)$. The quantity $\langle f(t)f(0)\rangle$
is plotted in Fig. \ref{fdt}. The figure is in good agreement with our
expectation that $\mu(t)\sim t^{-(\gamma/\nu-1)/z_c}$; this result has
also been observed for the the 2D Ising model with Glauber dynamics
\cite{zhong1}.

\section{Conclusion \label{Sec5}}

In this paper, we have studied the Fourier modes of the
two\text{-}dimensional Ising model with Kawasaki dynamics at critical
temperature and at zero (conserved) order parameter. We have
established that the Fourier modes are the dynamical eigenmodes of the
system to a very good approximation. Using these modes, we can
reconstruct the dynamics of any dynamical variable; we have done so
for the autocorrelation function and the mean-square deviation
(MSD) of line magnetization.

At the intermediate times, we have found that for $1\lesssim t
\lesssim L^{z_c}$, the line magnetisation undergoes anomalous
diffusion. We have argued that like other spin models and polymeric
systems this anomalous behavior can be described by the GLE
formulation with a memory kernel. The corresponding
fluctuation-dissipation theorem has been verified by the calculation
of the force autocorrelation.

With these results, we have showcased that for Kawasaki dynamics, the
Fourier modes, as the approximate dynamical eigenmodes, is a useful
tool to analytically derive the dynamical quantities in the Ising
system. We however note that if the model is evolved using Glauber
dynamics, then we find that $X_{pp}(t)$ decays as a stretched
exponential in time (not shown in this paper), which clearly shows
that the Fourier modes are not the (approximate) dynamical
eigenmodes. We do not understand this at present. It could be
explored in the future.

\section*{Acknowlegement \label{sec5}}

We thank R. C. Ball for valuable discussions. W.Z. acknowledges
financial support from the China Scholarship Council (CSC).

\section*{Appendix: Scaling of $\langle|A_p|\rangle^2$ with $p$ for the 2D Ising model\label{sec_apd}}

\setcounter{figure}{0} \renewcommand{\thefigure}{A\arabic{figure}}
\setcounter{equation}{0} \renewcommand{\theequation}{A\arabic{equation}}

In this appendix we obtain the scaling behaviour of
$\langle|A_p|^2\rangle$ for the 2D Ising model (note that the
calculations presented here do not correspond to the total
magnetisation of the sample kept fixed at zero, as is the case for
Kawasaki dynamics in this paper).

\begin{figure*}
\includegraphics[width=0.3\linewidth]{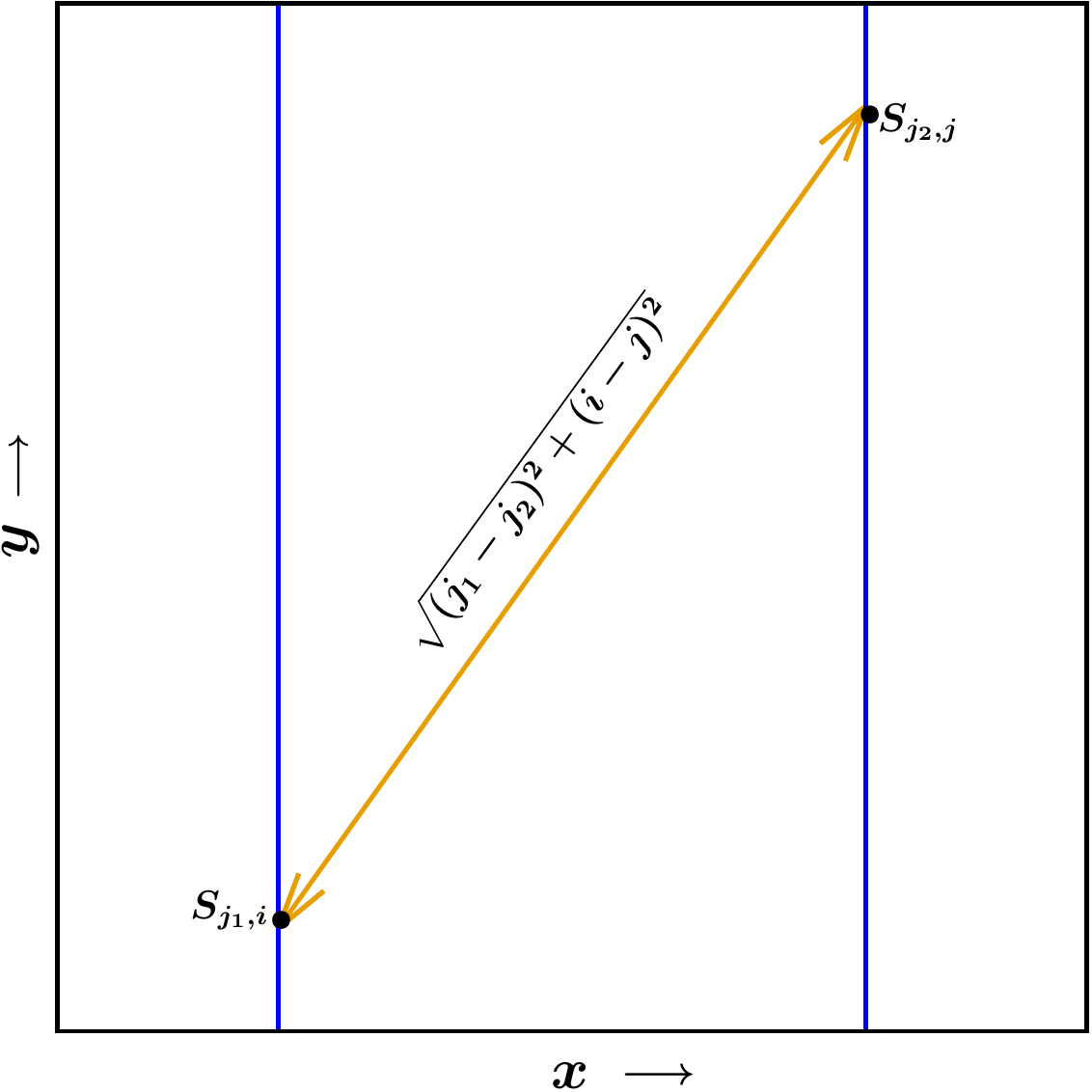} 
  \caption{Schematic diagram for the calculation of the line-line
    autocorrelation function.}
  \label{apen}
\end{figure*} 
 
First we calculate the autocorrelation function of the line
magnetisation. We use the classic result that at the critical
temperature the spin-spin autocorrelation function decays as
$r^{-\eta}$, where $r$ is the Euclidean distance between the two spins
and $\eta=2-\gamma/\nu=0.25$ for the 2D Ising model. With that
knowledge, upon summing over $i$ and $j$ in the $y$-direction (see
Fig. \ref{apen}), we obtain
\begin{equation}
\begin{aligned}
\langle M_l(j_1,0)M_l(j_2,0)\rangle =\sum_{i=0}^{L-1}\sum_{j=0}^{L-1}
\langle s_{j_1,i}\, s_{j_2,j} \rangle\sim \sum_{i=0}^{L-1}
\sum_{j=0}^{L-1}[(i-j)^2+(j_2-j_1)^2]^{-\eta/2}.
\end{aligned}
\label{ea1}
\end{equation}
We next set $a=(j_1-j_2)/L$, $u=(i-j)/L$ and $v=j/L$ to write,
\begin{equation}
\langle M_l(j_1,0)M_l(j_2,0)\rangle 
\sim\int_{-1}^1 \!\!du\,\,\frac{L^{2-\eta}}{[u^2+4a^2]^{\eta/2}}.
\label{ea2}
\end{equation}

The calculation of $\langle|A_p|^2\rangle$ follows from
Eq. (\ref{ea2}) in a similar manner.
\begin{equation}
\begin{aligned}
\langle|A_p|^2\rangle
=& \frac{1}{L^2}\sum_{j_1=0}^{L}\sum_{j_2=0}^{L} \langle M_l(j_1,0) M_l(j_2,0) \rangle \cos[2 \pi p (j_1-j_2)/L] .
\end{aligned}
\label{ea3}
\end{equation}
This time setting $a\rightarrow a/2$, Eq. (\ref{ea3}) reduces to
\begin{equation}
\begin{aligned}
\langle|A_p|^2\rangle\sim L^{2-\eta}\int_{-1}^1\!\!da \int_{-1}^1 \!\!du\,\,\frac{1}{[u^2+a^2]^{\eta/2}} \cos(\pi pa)=4L^{2-\eta}\int_0^1\!\!da \int_0^1 \!\!du\,\,\frac{1}{[u^2+a^2]^{\eta/2}} \cos(\pi pa).
\end{aligned}
\label{ea4}
\end{equation}
For $p=0$, Eq. (\ref{ea4}) leads to $|A_p(0)| ^2\sim L^{2-\eta}$,
which is the classic result for the equilibrium scaling $\langle
M^2\rangle\sim L^{4-\eta}=L^{2+\gamma/\nu}$ for the total sample
magnetisation $M$ for the 2D Ising model.

For $p\neq0$ we perform the integration over $u$ in Eq. (\ref{ea4}) to obtain
\begin{equation}
\begin{aligned}
\langle|A_p|^2\rangle\sim L^{2-\eta}\underbrace{\int_0^1\!\!da\,f(a)\cos(\pi pa)}_{I(p)},
\end{aligned}
\label{ea5}
\end{equation}
with
\begin{equation}
\begin{aligned}
  f(a)=\frac{(1\!+\!a^2)^{1-\eta/2}(5\!+\!a^2\!-\!\eta)}{(4\!-\!\eta)(2\!-\!\eta)}\!-\!\frac{(1\!+\!a^2)^{3-\eta/2}\,\,\text{Hypergeometric}_2F_1\left(1,(3\!-\!\eta)/2,-1/2,-1/a^2\right)}{a^2(4\!-\!\eta)(2\!-\!\eta)}.
\end{aligned}
\label{ea6}
\end{equation}
\begin{figure}[h]
\includegraphics[width=0.4\linewidth]{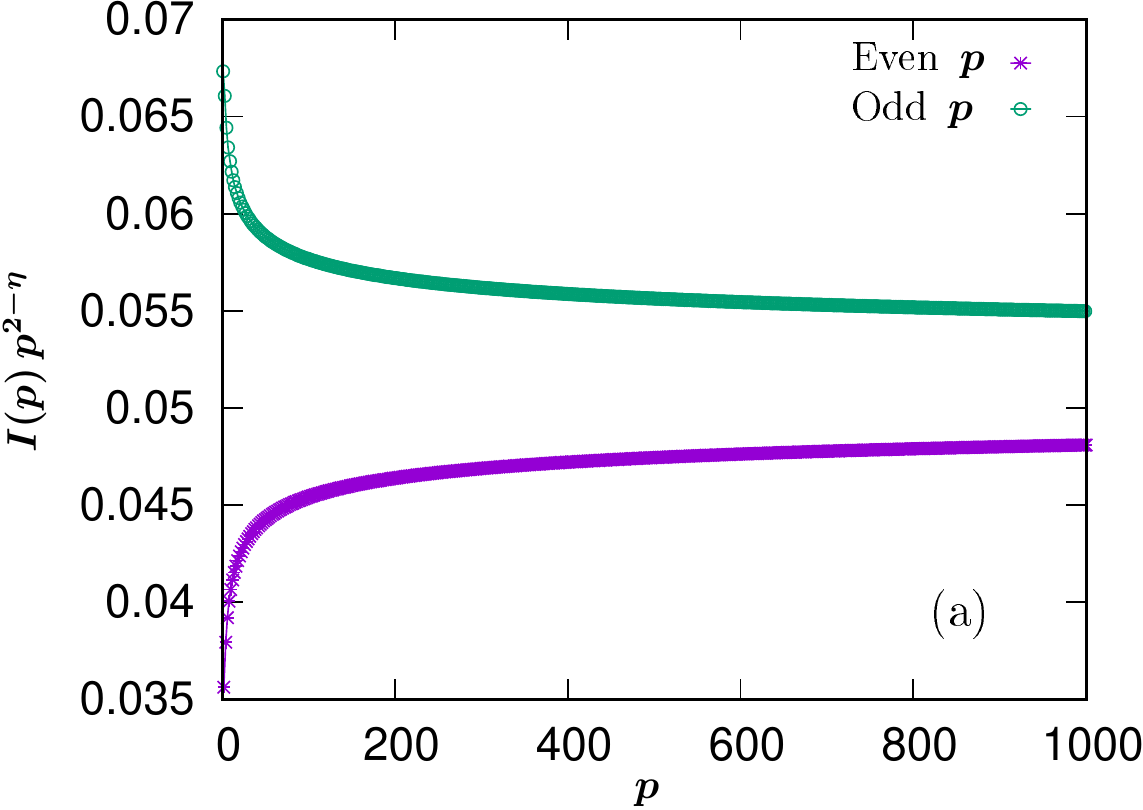}\hspace{1cm}\includegraphics[width=0.4\linewidth]{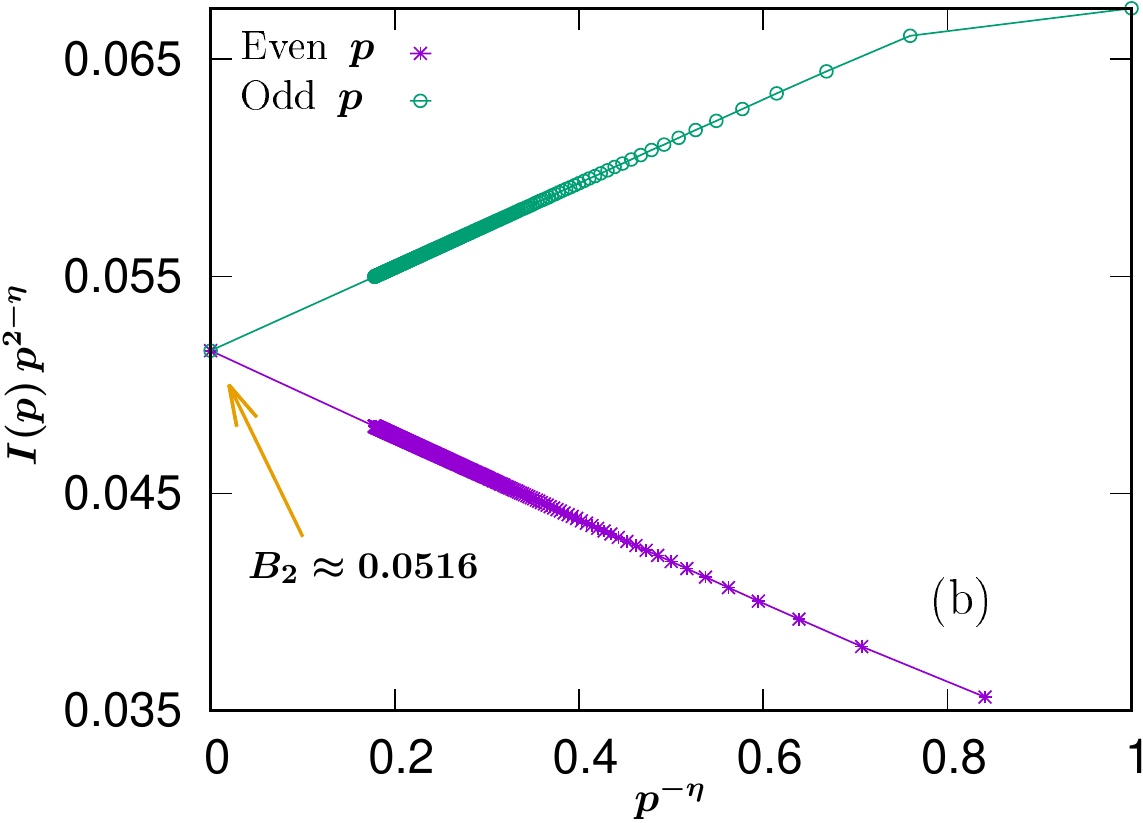}
  \caption{(a) Numerical integration of the integral in
    Eq. (\ref{ea6}). (b) The figure shows that
    $I(p)\approx B_2\,p^{2-\eta}$ in the limit $p\rightarrow\infty$, where $B_2\approx 0.0516$,
    although convergence to the asymptotic behaviour is rather
    slow. The solid lines are fits to the data.}
  \label{apen2}
\end{figure}

We then perform numerical integration separately for even and odd
$p$-values for Eq. (\ref{ea6}). The results, shown in
Fig. \ref{apen2}, demonstrate that in the limit $p\rightarrow\infty$
\begin{equation}
\begin{aligned}
\langle|A_p|\rangle^2\sim \left(\frac Lp\right)^{2-\eta}=\left(\frac Lp\right)^{\gamma/\nu},
\end{aligned}
\label{ea9}
\end{equation}
although convergence to the asymptotic behaviour is rather slow.

\end{document}